\documentclass{an}
\usepackage{graphicx}
\usepackage{times}
\begin{document}

\Pagespan{1}{}
\Yearpublication{\dots}
\Yearsubmission{\dots}
\Month{\dots}
\Volume{\dots}
\Issue{\dots}
\DOI{This.is/not.aDOI}

\title{Johnson-Cousins magnitudes of comparison stars in the fields of ten Seyfert
galaxies\thanks{Based on observations obtained with the 2-m telescope of the Rozhen National
Astronomical Observatory, which is operated by the Institute of Astronomy, Bulgarian Academy of Sciences.}}

\author{B. M. Mihov\thanks{Corresponding author: \email{bmihov@astro.bas.bg}} \and L. S. Slavcheva-Mihova}

\titlerunning{$UBVR_{\rm \scriptstyle C}I_{\rm \scriptstyle C}$ magnitudes of stars in the fields of ten Seyferts}
\authorrunning{B. M. Mihov \& L. S. Slavcheva-Mihova}

\institute{Institute of Astronomy, Bulgarian Academy of Sciences, 72 Tsarigradsko Chausse
Blvd., 1784 Sofia, Bulgaria}

\received{\ldots}
\accepted{\ldots}
\publonline{later}


\abstract{We present $UBVR_{\rm \scriptstyle C}I_{\rm \scriptstyle C}$ magnitudes
of 49 comparison stars in the fields of the Seyfert galaxies Mrk~335, Mrk~79, Mrk~279, Mrk~506,
3C~382, 3C~390.3, NGC~6814, Mrk~304, Ark~564, and NGC~7469 in order to facilitate the photometric
monitoring of these objects; 36 of the stars have not been calibrated before. The comparison
stars are situated in $5 \times 5$ arcmin fields centred on the Seyfert galaxies, their $V$
band flux ranges from 11.7 to 18.2 mag with a median value of 16.3 mag, and their $B-V$ colour
index ranges from 0.4 to 1.6 mag with a median value of 0.8 mag. The median errors of the
calibrated $UBVR_{\rm \scriptstyle C}I_{\rm \scriptstyle C}$ magnitudes are 0.08,
0.04, 0.03, 0.04, and 0.06 mag, respectively. Comparison stars were calibrated for the first
time in three of the fields (Mrk~506, 3C~382, and Mrk~304). The comparison sequences in the other
fields were improved in various aspects. Extra stars were calibrated in four fields (Mrk~335,
Mrk~79, NGC~6814, and NGC~7469)~-- most of these stars are fainter and are situated closer to
the Seyfert galaxies compared to the existing comparison stars. The passband coverage of the
sequences in five fields (Mrk~335, Mrk~79, Mrk~279, NGC~6814, and Ark~564) was complemented
with $U$ band.}

\maketitle

\section{Introduction}
\label{intro}

Variability is a common property of Active Galactic Nuclei (AGNs). Monitoring programmes
undertaken till now provided a lot of information about variability characteristics
of AGNs that was successfully used in studying the nuclear activity of galaxies (e.g.
Schramm et~al. \cite{schramm93}; Dietrich et~al. \cite{dietrich98}; Shemmer et~al.
\cite{shemmer01}; Bachev \& Strigachev \cite{bachev04}; Bachev, Strigachev \& Semkov
\cite{bachev05}; Villata et~al. \cite{villata06}).

\begin{table*}[t]
\caption{A list of the selected Seyfert galaxies. Equatorial coordinates, other names, redshifts, and
Seyfert types of activity were taken from NASA/IPAC Extragalactic Database (NED). The number of the
different observing nights, $N$, at which the data were taken is specified in the last column; the
$U$ band data of Mrk~335, Mrk~79, Mrk~506, NGC~6814, and Mrk~304 were taken at a single epoch. The
superscript $n$ means the number of nights used in the internal check for variability of the newly
calibrated stars (see Sect.~\ref{genchar}); the missing superscript means no new comparison stars
were added to the corresponding fields.}
\label{gals}
\centering
\begin{tabular}{@{}llrrrrl@{}}
\noalign{\smallskip} \hline
\noalign{\smallskip}
$\rm Galaxy$ & $\rm Other$ $\rm names$ & $\rm RA\,(J2000)$ & $\rm Dec\,(J2000)$ & $z$ & $\rm Sy$ & $N^n$ \\
\noalign{\smallskip} \hline
\noalign{\smallskip}
$\rm Mrk~335  $ &                                                    & $00$:$06$:$19.521$ & $ 20$:$12$:$10.49$ & $0.02578$ & $1.2$ & $2^3$ \\
$\rm Mrk~79   $ & $\rm UGC\,03973,\,CGCG\,235-030,\,MCG\,+08-14-033$ & $07$:$42$:$32.797$ & $ 49$:$48$:$34.75$ & $0.02219$ & $1.2$ & $2^3$ \\
$\rm Mrk~279  $ & $\rm UGC\,08823,\,CGCG\,336-028,\,MCG\,+12-13-022$ & $13$:$53$:$03.447$ & $ 69$:$18$:$29.57$ & $0.03045$ & $1.5$ & $1  $ \\
$\rm Mrk~506  $ & $\rm CGCG\,170-020,\,MCG\,+05-41-012             $ & $17$:$22$:$39.899$ & $ 30$:$52$:$53.01$ & $0.04303$ & $1.5$ & $2^3$ \\
$\rm 3C~382   $ & $\rm CGCG\,173-014                               $ & $18$:$35$:$03.391$ & $ 32$:$41$:$46.82$ & $0.05787$ & $1.0$ & $2^3$ \\
$\rm 3C~390.3 $ & $\rm VII\,Zw\,838                                $ & $18$:$42$:$08.990$ & $ 79$:$46$:$17.13$ & $0.05610$ & $1.0$ & $2  $ \\
$\rm NGC~6814 $ & $\rm MCG\,-02-50-001                             $ & $19$:$42$:$40.644$ & $-10$:$19$:$24.57$ & $0.00521$ & $1.5$ & $3^4$ \\
$\rm Mrk~304  $ & $\rm II\,Zw\,175,\,CGCG\,428-065                 $ & $22$:$17$:$12.260$ & $ 14$:$14$:$20.90$ & $0.06576$ & $1.0$ & $2^4$ \\
$\rm Ark~564  $ & $\rm UGC\,12163,\,CGCG\,495-018,\,MCG\,+05-53-012$ & $22$:$42$:$39.345$ & $ 29$:$43$:$31.31$ & $0.02468$ & $1.8$ & $2  $ \\
$\rm NGC~7469 $ & $\rm UGC\,12332,\,CGCG\,405-026,\,MCG\,+01-58-025$ & $23$:$03$:$15.623$ & $ 08$:$52$:$26.39$ & $0.01632$ & $1.2$ & $2^4$ \\
\noalign{\smallskip} \hline
\noalign{\smallskip}
\end{tabular}
\end{table*}

\begin{table*}
\caption{Apparent Johnson-Cousins $UBVR_{\rm \scriptstyle C}I_{\rm \scriptstyle C}$
magnitudes of the comparison stars, their errors (in parentheses), and the number
of the individual estimates used to obtain the final magnitude. We list the other
designations of the comparison stars that have been calibrated by other authors as
well; the superscripts to these designations refer to the corresponding literature
sources. The stars without other designations have been calibrated for the first
time by us. Dots mean we have not been able to obtain the corresponding magnitudes
(see Sect.~\ref{comm} for comments).}
\label{naomag}
\centering
\begin{tabular}{@{}l@{\hspace{3.2mm}}r@{\hspace{3.2mm}}r@{\hspace{3.2mm}}r@{\hspace{3.2mm}}r@{\hspace{3.2mm}}r@{\hspace{3.2mm}}r@{\hspace{3.2mm}}r@{}}
\noalign{\smallskip} \hline
\noalign{\smallskip}
$\rm Star$ & $\rm RA\,(J2000)$ & $\rm Dec\,(J2000)$ & $U$ & $B$ & $V$ & $R_{\rm \scriptstyle C}$ & $I_{\rm \scriptstyle C}$ \\
\noalign{\smallskip} \hline
\noalign{\smallskip}
\multicolumn{8}{c}{$\rm \textbf {Mrk~335}$} \\
$\rm \textbf {A}$:$\;\rm 1^{\rm b},\,B^{\rm c},\,4^{\rm e} $ & $00$:$06$:$20.107$ & $20$:$10$:$50.59$ & $16.17\;(0.08)\;1$ & $15.42\;(0.03)\;2$ & $14.25\;(0.03)\;2$ & $13.70\;(0.03)\;2$ & $13.12\;(0.06)\;2$ \\
$\rm \textbf {B}                                           $ & $00$:$06$:$14.808$ & $20$:$11$:$34.20$ & $17.31\;(0.08)\;1$ & $17.38\;(0.03)\;2$ & $16.85\;(0.03)\;2$ & $16.55\;(0.03)\;2$ & $16.22\;(0.05)\;2$ \\
$\rm \textbf {C}$:$\;\rm 2^{\rm b},\,C^{\rm c},\,6^{\rm e} $ & $00$:$06$:$17.971$ & $20$:$13$:$17.60$ & $15.40\;(0.08)\;1$ & $15.43\;(0.03)\;2$ & $15.00\;(0.03)\;2$ & $14.78\;(0.03)\;2$ & $14.55\;(0.06)\;2$ \\
$\rm \textbf {D}                                           $ & $00$:$06$:$17.135$ & $20$:$14$:$08.72$ & $16.28\;(0.09)\;1$ & $16.14\;(0.04)\;2$ & $15.38\;(0.03)\;2$ & $15.00\;(0.04)\;2$ & $14.59\;(0.06)\;2$ \\
\noalign{\smallskip} \hline
\noalign{\smallskip}
\multicolumn{8}{c}{$\rm \textbf {Mrk~79}$} \\
$\rm \textbf {A}                                     $ & $07$:$42$:$35.026$ & $49$:$47$:$01.59$ & $19.18\;(0.09)\;1$ & $18.84\;(0.04)\;2$ & $17.94\;(0.06)\;2$ & $17.48\;(0.05)\;2$ & $17.05\;(0.11)\;2$ \\
$\rm \textbf {B}                                     $ & $07$:$42$:$25.704$ & $49$:$48$:$07.52$ & $17.62\;(0.09)\;1$ & $17.60\;(0.05)\;2$ & $16.88\;(0.05)\;2$ & $16.53\;(0.05)\;2$ & $16.13\;(0.08)\;2$ \\
$\rm \textbf {C}$:$\;4^{\rm e}                       $ & $07$:$42$:$22.633$ & $49$:$48$:$17.49$ & $15.02\;(0.10)\;1$ & $14.98\;(0.06)\;1$ & $14.30\;(0.05)\;1$ & $13.98\;(0.06)\;1$ & $13.61\;(0.07)\;1$ \\
$\rm \textbf {D}                                     $ & $07$:$42$:$42.113$ & $49$:$49$:$46.54$ & $20.08\;(0.09)\;1$ & $18.66\;(0.04)\;2$ & $17.42\;(0.06)\;2$ & $16.70\;(0.07)\;2$ & $16.04\;(0.07)\;2$ \\
$\rm \textbf {E}                                     $ & $07$:$42$:$24.547$ & $49$:$49$:$54.38$ & $19.46\;(0.09)\;1$ & $18.39\;(0.08)\;2$ & $17.01\;(0.04)\;2$ & $16.12\;(0.08)\;2$ & $15.22\;(0.06)\;2$ \\
\noalign{\smallskip} \hline
\noalign{\smallskip}
\multicolumn{8}{c}{$\rm \textbf {Mrk~279}$} \\
$\rm \textbf {A}$:$\;\rm D^{\rm c,f}                 $ & $13$:$53$:$21.781$ & $69$:$16$:$30.34$ & $16.37\;(0.06)\;1$ & $16.23\;(0.05)\;1$ & $15.53\;(0.03)\;1$ & $15.16\;(0.05)\;1$ & $14.82\;(0.08)\;1$ \\
$\rm \textbf {B}$:$\;\rm A^{\rm c,f}                 $ & $13$:$53$:$21.294$ & $69$:$18$:$07.97$ & $14.07\;(0.05)\;1$ & $13.11\;(0.05)\;1$ & $12.06\;(0.03)\;1$ & $          \ldots$ & $          \ldots$ \\
$\rm \textbf {C}$:$\;\rm C^{\rm c,f}                 $ & $13$:$52$:$54.690$ & $69$:$20$:$14.75$ & $16.69\;(0.07)\;1$ & $15.79\;(0.05)\;1$ & $14.78\;(0.03)\;1$ & $14.23\;(0.06)\;1$ & $13.81\;(0.09)\;1$ \\
\noalign{\smallskip} \hline
\noalign{\smallskip}
\multicolumn{8}{c}{$\rm \textbf {Mrk~506}$} \\
$\rm \textbf {A}                                     $ & $17$:$22$:$37.829$ & $30$:$50$:$10.71$ & $20.01\;(0.06)\;1$ & $19.13\;(0.03)\;2$ & $17.72\;(0.02)\;2$ & $16.90\;(0.04)\;2$ & $16.03\;(0.05)\;2$ \\
$\rm \textbf {B}                                     $ & $17$:$22$:$32.619$ & $30$:$51$:$23.33$ & $19.02\;(0.05)\;1$ & $18.23\;(0.03)\;2$ & $17.12\;(0.02)\;2$ & $16.49\;(0.04)\;2$ & $15.87\;(0.04)\;2$ \\
$\rm \textbf {C}                                     $ & $17$:$22$:$43.293$ & $30$:$52$:$00.91$ & $20.65\;(0.05)\;1$ & $19.49\;(0.03)\;2$ & $17.92\;(0.02)\;2$ & $16.86\;(0.04)\;2$ & $15.45\;(0.04)\;2$ \\
$\rm \textbf {D}                                     $ & $17$:$22$:$33.955$ & $30$:$52$:$02.37$ & $17.85\;(0.05)\;1$ & $16.62\;(0.03)\;2$ & $15.33\;(0.02)\;2$ & $14.59\;(0.04)\;2$ & $13.92\;(0.04)\;2$ \\
$\rm \textbf {E}                                     $ & $17$:$22$:$47.853$ & $30$:$52$:$12.09$ & $17.86\;(0.06)\;1$ & $18.03\;(0.03)\;2$ & $17.46\;(0.02)\;2$ & $17.14\;(0.05)\;2$ & $16.80\;(0.05)\;2$ \\
$\rm \textbf {F}                                     $ & $17$:$22$:$42.842$ & $30$:$52$:$59.42$ & $14.92\;(0.05)\;1$ & $15.05\;(0.03)\;2$ & $14.61\;(0.02)\;2$ & $14.38\;(0.03)\;2$ & $14.13\;(0.04)\;2$ \\
\noalign{\smallskip} \hline
\noalign{\smallskip}
\multicolumn{8}{c}{$\rm \textbf {3C~382}$} \\
$\rm \textbf {A}                                     $ & $18$:$35$:$06.547$ & $32$:$40$:$02.03$ & $18.32\;(0.09)\;2$ & $17.46\;(0.05)\;1$ & $16.42\;(0.03)\;2$ & $15.82\;(0.04)\;2$ & $15.37\;(0.05)\;2$ \\
$\rm \textbf {B}                                     $ & $18$:$34$:$57.385$ & $32$:$41$:$17.26$ & $17.37\;(0.12)\;2$ & $17.00\;(0.05)\;1$ & $16.27\;(0.02)\;2$ & $15.86\;(0.03)\;2$ & $15.57\;(0.04)\;2$ \\
$\rm \textbf {C}                                     $ & $18$:$35$:$01.184$ & $32$:$42$:$43.29$ & $17.86\;(0.07)\;2$ & $17.99\;(0.05)\;1$ & $17.41\;(0.02)\;2$ & $17.03\;(0.02)\;2$ & $16.77\;(0.05)\;2$ \\
$\rm \textbf {D}                                     $ & $18$:$35$:$05.136$ & $32$:$42$:$44.90$ & $17.72\;(0.21)\;2$ & $16.80\;(0.05)\;1$ & $15.76\;(0.03)\;2$ & $15.18\;(0.02)\;2$ & $14.75\;(0.04)\;2$ \\
$\rm \textbf {E}                                     $ & $18$:$35$:$01.718$ & $32$:$42$:$58.12$ & $17.34\;(0.13)\;2$ & $17.16\;(0.05)\;1$ & $16.51\;(0.02)\;2$ & $16.10\;(0.02)\;2$ & $15.80\;(0.04)\;2$ \\
$\rm \textbf {F}                                     $ & $18$:$35$:$07.086$ & $32$:$43$:$27.46$ & $16.30\;(0.12)\;2$ & $15.83\;(0.05)\;1$ & $15.00\;(0.02)\;2$ & $14.52\;(0.02)\;2$ & $14.16\;(0.04)\;2$ \\
$\rm \textbf {G}                                     $ & $18$:$34$:$57.009$ & $32$:$43$:$33.51$ & $15.95\;(0.04)\;2$ & $15.99\;(0.05)\;1$ & $15.55\;(0.02)\;2$ & $15.15\;(0.02)\;2$ & $14.87\;(0.05)\;2$ \\
\noalign{\smallskip} \hline
\noalign{\smallskip}
\multicolumn{8}{c}{$\rm \textbf {3C~390.3}$} \\
$\rm \textbf {A}$:$\;\rm A^{\rm a},\,1^{\rm b}        $ & $18$:$42$:$23.706$ & $79$:$45$:$39.09$ & $13.65\;(0.08)\;2$ & $12.74\;(0.04)\;2$ & $11.73\;(0.08)\;2$ & $11.11\;(0.05)\;1$ & $10.66\;(0.07)\;1$ \\
$\rm \textbf {B}$:$\;\rm B^{\rm a,f},\,2^{\rm b}      $ & $18$:$42$:$00.991$ & $79$:$47$:$35.25$ & $15.12\;(0.05)\;2$ & $15.02\;(0.04)\;2$ & $14.31\;(0.03)\;2$ & $13.94\;(0.06)\;2$ & $13.56\;(0.05)\;2$ \\
$\rm \textbf {C}$:$\;\rm D^{\rm a,f},\,3^{\rm b}      $ & $18$:$41$:$29.706$ & $79$:$47$:$59.64$ & $15.72\;(0.06)\;2$ & $15.45\;(0.05)\;2$ & $14.66\;(0.05)\;2$ & $14.29\;(0.08)\;2$ & $13.92\;(0.07)\;2$ \\
\noalign{\smallskip} \hline
\noalign{\smallskip}
\multicolumn{8}{c}{$\rm \textbf {NGC~6814}$} \\
$\rm \textbf {A}                                     $ & $19$:$42$:$34.053$ & $-10$:$21$:$03.98$ & $          \ldots$ & $19.43\;(0.08)\;2$ & $17.91\;(0.02)\;3$ & $17.16\;(0.03)\;3$ & $16.47\;(0.06)\;2$ \\
$\rm \textbf {B}                                     $ & $19$:$42$:$32.218$ & $-10$:$18$:$55.05$ & $17.75\;(0.08)\;1$ & $17.28\;(0.03)\;2$ & $16.31\;(0.02)\;3$ & $15.84\;(0.03)\;3$ & $15.40\;(0.05)\;2$ \\
$\rm \textbf {C}                                     $ & $19$:$42$:$38.269$ & $-10$:$17$:$46.90$ & $16.71\;(0.06)\;1$ & $15.98\;(0.03)\;2$ & $14.90\;(0.02)\;3$ & $14.40\;(0.03)\;3$ & $13.91\;(0.04)\;2$ \\
$\rm \textbf {D}                                     $ & $19$:$42$:$47.459$ & $-10$:$17$:$42.35$ & $18.06\;(0.08)\;1$ & $17.42\;(0.03)\;2$ & $16.38\;(0.02)\;3$ & $15.88\;(0.03)\;3$ & $15.41\;(0.05)\;2$ \\
$\rm \textbf {E}                                     $ & $19$:$42$:$45.795$ & $-10$:$17$:$40.85$ & $17.81\;(0.07)\;1$ & $16.33\;(0.03)\;2$ & $14.93\;(0.03)\;3$ & $14.28\;(0.03)\;3$ & $13.68\;(0.05)\;2$ \\
$\rm \textbf {F}                                     $ & $19$:$42$:$36.006$ & $-10$:$17$:$22.74$ & $17.52\;(0.07)\;1$ & $16.92\;(0.03)\;2$ & $15.93\;(0.02)\;3$ & $15.45\;(0.03)\;3$ & $15.01\;(0.05)\;2$ \\
\noalign{\smallskip} \hline
\noalign{\smallskip}
\multicolumn{8}{c}{$\rm \textbf {Mrk~304}$} \\
$\rm \textbf {A}                                     $ & $22$:$17$:$17.819$ & $14$:$13$:$28.11$ & $15.69\;(0.06)\;1$ & $15.42\;(0.04)\;2$ & $14.65\;(0.03)\;2$ & $14.23\;(0.04)\;2$ & $13.84\;(0.06)\;2$ \\
$\rm \textbf {B}                                     $ & $22$:$17$:$04.362$ & $14$:$14$:$37.28$ & $17.86\;(0.06)\;1$ & $17.82\;(0.04)\;2$ & $17.09\;(0.03)\;2$ & $16.68\;(0.04)\;2$ & $16.31\;(0.06)\;2$ \\
$\rm \textbf {C}                                     $ & $22$:$17$:$14.323$ & $14$:$15$:$08.79$ & $17.58\;(0.05)\;1$ & $17.66\;(0.04)\;2$ & $16.98\;(0.03)\;2$ & $16.55\;(0.04)\;2$ & $16.15\;(0.06)\;2$ \\
$\rm \textbf {D}                                     $ & $22$:$17$:$08.602$ & $14$:$15$:$25.62$ & $19.32\;(0.05)\;1$ & $18.03\;(0.04)\;2$ & $16.85\;(0.03)\;2$ & $16.16\;(0.04)\;2$ & $15.59\;(0.06)\;2$ \\
$\rm \textbf {E}                                     $ & $22$:$17$:$05.070$ & $14$:$15$:$29.10$ & $17.02\;(0.06)\;1$ & $16.31\;(0.04)\;2$ & $15.33\;(0.03)\;2$ & $14.78\;(0.04)\;2$ & $14.27\;(0.06)\;2$ \\
$\rm \textbf {F}                                     $ & $22$:$17$:$20.088$ & $14$:$15$:$43.74$ & $18.47\;(0.07)\;1$ & $18.32\;(0.05)\;2$ & $17.59\;(0.03)\;2$ & $17.11\;(0.05)\;2$ & $16.70\;(0.07)\;2$ \\
$\rm \textbf {G}                                     $ & $22$:$17$:$19.755$ & $14$:$16$:$06.40$ & $17.44\;(0.07)\;1$ & $17.38\;(0.05)\;2$ & $16.65\;(0.03)\;2$ & $16.21\;(0.05)\;2$ & $15.81\;(0.07)\;2$ \\
\noalign{\smallskip} \hline
\noalign{\smallskip}
\end{tabular}
\end{table*}

\setcounter{table}{1}

\begin{table*}[t]
\caption{Continued.}
\centering
\begin{tabular}{@{}l@{\hspace{4.0mm}}r@{\hspace{4.0mm}}r@{\hspace{4.0mm}}r@{\hspace{4.0mm}}r@{\hspace{4.0mm}}r@{\hspace{4.0mm}}r@{\hspace{4.0mm}}r@{}}
\noalign{\smallskip} \hline
\noalign{\smallskip}
$\rm Star$ & $\rm RA\,(J2000)$ & $\rm Dec\,(J2000)$ & $U$ & $B$ & $V$ & $R_{\rm \scriptstyle C}$ & $I_{\rm \scriptstyle C}$ \\
\noalign{\smallskip} \hline
\noalign{\smallskip}
\multicolumn{8}{c}{$\rm \textbf {Ark~564}$} \\
$\rm \textbf {A}$:$\;3^{\rm d}                       $ & $22$:$42$:$39.265$ & $29$:$44$:$21.03$ & $14.16\;(0.09)\;2$ & $14.17\;(0.07)\;2$ & $13.58\;(0.05)\;2$ & $13.27\;(0.07)\;2$ & $12.97\;(0.06)\;2$ \\
$\rm \textbf {B}$:$\;1^{\rm f}                       $ & $22$:$42$:$48.379$ & $29$:$45$:$22.30$ & $13.90\;(0.13)\;2$ & $13.97\;(0.09)\;2$ & $13.44\;(0.07)\;2$ & $13.17\;(0.09)\;2$ & $12.93\;(0.08)\;2$ \\
$\rm \textbf {C}$:$\;2^{\rm d},\,13^{\rm f}          $ & $22$:$42$:$32.084$ & $29$:$45$:$27.02$ & $15.85\;(0.11)\;2$ & $15.77\;(0.08)\;2$ & $15.07\;(0.07)\;2$ & $14.74\;(0.08)\;2$ & $14.43\;(0.08)\;2$ \\
$\rm \textbf {D}$:$\;1^{\rm d},\,12^{\rm f}          $ & $22$:$42$:$35.122$ & $29$:$45$:$37.96$ & $15.56\;(0.09)\;2$ & $15.42\;(0.08)\;2$ & $14.73\;(0.07)\;2$ & $14.41\;(0.09)\;2$ & $14.12\;(0.08)\;2$ \\
\noalign{\smallskip} \hline
\noalign{\smallskip}
\multicolumn{8}{c}{$\rm \textbf {NGC~7469}$} \\
$\rm \textbf {A}                                     $ & $23$:$03$:$10.785$ & $08$:$50$:$43.14$ & $18.59\;(0.17)\;2$ & $18.81\;(0.05)\;2$ & $18.20\;(0.04)\;2$ & $17.93\;(0.04)\;2$ & $17.55\;(0.06)\;2$ \\
$\rm \textbf {B}                                     $ & $23$:$03$:$07.717$ & $08$:$51$:$43.43$ & $17.22\;(0.10)\;2$ & $17.38\;(0.05)\;2$ & $16.73\;(0.05)\;2$ & $16.43\;(0.04)\;2$ & $16.02\;(0.06)\;2$ \\
$\rm \textbf {C}                                     $ & $23$:$03$:$12.216$ & $08$:$52$:$12.93$ & $17.46\;(0.10)\;2$ & $17.65\;(0.04)\;2$ & $17.06\;(0.04)\;2$ & $16.75\;(0.04)\;2$ & $16.32\;(0.05)\;2$ \\
$\rm \textbf {D}                                     $ & $23$:$03$:$23.233$ & $08$:$53$:$21.83$ & $18.34\;(0.13)\;2$ & $18.62\;(0.06)\;2$ & $17.96\;(0.05)\;2$ & $17.67\;(0.04)\;2$ & $17.29\;(0.06)\;2$ \\
\noalign{\smallskip} \hline
\noalign{\smallskip}
\end{tabular}
\flushleft{$^{\rm a}$\,Penston et~al. (\cite{penston71}), $^{\rm b}$\,Curry et~al. (\cite{curry98}),
$^{\rm c}$\,Bachev et~al. (\cite{bachev00}), $^{\rm d}$\,Shemmer et~al. (\cite{shemmer01}),
$^{\rm e,\,f}$\,Doroshenko et~al. (\cite{doroshenko05a}, \cite{doroshenko05b}).}
\end{table*}

Calibration of optical monitoring light curves is important as AGN monitoring programmes
typically combine data from different observatories~-- to get dense temporal coverage
of the resulting light curves over a certain period of time~-- and from different
frequency domains~-- to study the AGN emission across the electromagnetic spectrum.
The preferable photometric technique used to calibrate AGN light curves is
relative photometry against local comparison stars. Compared to multi-airmass calibration,
it is less time-consuming and makes possible the use of non-photometric nights, too.
Therefore, the presence of comparison sequences in the fields of target AGNs is of
importance for monitoring programmes. Aside from photometric monitoring of AGNs,
the comparison sequences could also be used for surface photometry calibration of the
corresponding active galaxies.

Johnson-Cousins comparison sequences in the fields of Seyfert galaxies were presented in
the papers of Penston, Penston \& Sandage (\cite{penston71}), Miller (\cite{miller81},\cite{miller86}),
Hamuy \& Maza (\cite{hamuy89}), Curry et~al. (\cite{curry98}), Bachev, Strigachev \& Dimitrov
(\cite{bachev00}), Gonz\'alez-P\'erez, Kidger \& Mart\'in-Luis (\cite{gonzalez01}), and Doroshenko
et~al. (\cite{doroshenko05a}, \cite{doroshenko05b}). Further improvement of the existing standard
sequences along with establishing new ones is highly recommended. The improvement of a comparison
star sequence could comprise:
\begin{itemize}
\item[--] Addition of new comparison stars in order to extend the magnitude and the colour
index ranges covered by the comparison sequence and/or to optimize the distribution
of the standards in the field of the target object. The larger number of comparison
stars would also increase the accuracy of the zero-point magnitude derived by them;
\item[--] Addition of new flux measurements for the existing comparison stars in order
to increase the accuracy of their magnitudes. Further check of the comparison stars
for variability could also be done;
\item[--] Extension of the passband coverage of the calibrated magnitudes.
\end{itemize}

Since 1997 $UBVR_{\rm \scriptstyle C}I_{\rm \scriptstyle C}$ observations
of a number of Seyfert galaxies have been obtained in the course of active galaxy surface
photometry programme at the Rozhen National Astronomical Observatory (NAO), Bulgaria,
by the authors; about a dozen of the objects have been observed two or more times. We
were not able to do a multi-airmass calibration of the data taken during a few of the
observing nights due to weather or technical problems. A part of these non-calibrated
data was transformed to the standard Johnson-Cousins system using comparison
sequences established in the corresponding object fields, whereas for the rest this
approach failed because of the following reasons: (1) the comparison stars lay
outside our field of view, (2) the standards were saturated in our frames, (3) the
standard sequence did not cover some of the photometric bands (usually $U$), and
(4) there was no comparison sequence established in the field. A possible approach
in these cases is matching the non-calibrated host galaxy profiles or galaxy
multi-aperture magnitudes to calibrated ones (e.g. Kotilainen, Ward \& Williger \cite{kotilainen93}).
However, this type of calibration is of lower accuracy compared to both multi-airmass
and relative calibration.

In order to transform the non-calibrated data of ours to the standard system
and to facilitate future AGN monitoring, we re-examined our observational data with the
purpose to find objects suitable (1) for establishing new comparison sequences
in their fields and (2) for improving the standard sequences already existing. A total
of ten Seyfert galaxies were selected as a result of this search; some information about
these objects is listed in Table~\ref{gals}. We present the results of the
$UBVR_{\rm \scriptstyle C}I_{\rm \scriptstyle C}$ calibration
of comparison stars in the fields of the selected galaxies in this paper.

The paper is organized as follows. The observations and data reduction are presented
in Sect.~\ref{obsred}. The flux measurement and magnitude calibration are described
in Sect.~\ref{fluxmc}. The general characteristics of the calibrated stars are discussed
in Sect.~\ref{genchar}. Comments on the individual fields are presented in Sect.~\ref{comm}
and we summarize the main results of this paper in Sect.~\ref{summ}.

\section{Observations and data reduction}
\label{obsred}

All observations were done by the authors using the 2-m telescope of NAO during
the period 1997~-- 2003. A liquid nitrogen cooled $1024 \times 1024$ Photometrics
AT200 model CCD camera (CCD chip SITe SI003AB having $24\,\mu\rm m$ square pixel)
was attached to the Ritchie-Chr\'etien focus of the telescope giving a scale
factor of $0.309\,\rm arcsec\,\rm px^{-1}$ and a square, 5.3 arcmin wide field
of view. A standard Johnson-Cousins $UBVR_{\rm \scriptstyle C}I_{\rm \scriptstyle C}$
set of filters was used. Multiple exposures of each object field were taken as a
rule. Zero exposure frames were taken regularly each observing run and flat fields
were taken during evening and/or morning twilights; dark current was negligible
as the CCD chip was cooled down to $-100$ degrees of Celsius. The typical size
of the seeing disk was about 2 arcsec. The binning factor of the CCD camera was
changed depending on the seeing conditions~-- we tried to ensure about 3 pixels
per FWHM in order to maximize the signal-to-noise ratio without loss of resolution.
Standard sequences established in stellar clusters were observed two or three times
each night in order to determine the transformation coefficients to the
standard Johnson-Cousins system. We used the clusters Messier~67 (Chevalier \& Ilovaisky
\cite{chevalier91}), Messier~92 (Majewski et~al. \cite{majewski94}), and NGC~7790
(Odewahn, Bryja \& Humphreys \cite{odewahn92}; Petrov et~al. \cite{petrov01}) for this purpose.
Note that we have added 0.002 mag to the $V$ magnitudes and to the $B-V$ colour indices
of the Messier~92 standard stars listed in Majewski et~al. (\cite{majewski94}) according
to the addendum of Stetson \& Harris (\cite{stetson88}). The cluster standards were
observed at airmass values between 1 and 2; in each case the programme fields were observed
within that range.

The initial processing of the frames was performed using $\rm \scriptstyle ESO-MIDAS$
package and following the standard CCD reduction steps. The mean bias level was estimated
using properly selected columns of the virtual prescan section of the CCD chip and then
subtracted. The removal of the residual bias pattern was done using a median zero exposure
frame, and the flat fielding and de-fringing were performed employing median flat field and
fringe frames, respectively; de-fringing was applied only to $I_{\rm \scriptstyle C}$
frames. Cosmic ray hits were cleaned up using $\rm \scriptstyle FILTER/COSMIC$ command.
The individual frames of each object in a given passband were aligned using
$\rm \scriptstyle ALIGN/IMAGE$ and $\rm \scriptstyle REBIN/ROTATE$ commands and then averaged.

\section{Flux measurement and calibration}
\label{fluxmc}

The flux measurements of all objects of interest were performed using $\rm \scriptstyle DAOPHOT$
package run within $\rm \scriptstyle ESO-MIDAS$ (Stetson \cite{stetson87}). The growth
curve technique was used to obtain the total instrumental magnitudes ensuring good accuracy
even for weak stars (Stetson \cite{stetson90}); we found no variations of PSF across
the CCD camera field and, therefore, the usage of this technique is justified\footnote{The
growth curve magnitudes of a number of comparison stars were found to be identical with the
single, $5 \times \rm FWHM$ radius aperture magnitudes to within the measurement errors. This
is another confirmation of the growth curve technique applicability in our measurements.}.
Neighbourhood objects to the stars used to build the mean growth curve of a given frame were
cleaned out using either the technique presented by Markov et~al. (\cite{markov97}) or the
star subtraction technique based on $\rm \scriptstyle DAOPHOT$. We were not able to derive
accurate aperture corrections for the $U$ band frames of 3C~382 and NGC~7469 due to the
lack of stars with high enough signal-to-noise ratio in our frames. We measured the
stars of interest directly in an aperture of $5 \times \rm FWHM$ radius in these
cases (usually at this radius the growth curve asymptotically flattens). The sky
background level was estimated as the mode of the pixel values in an annulus
having an inner radius of $7 \times \rm FWHM$ pixels and an outer radius
of $[\,1000/\pi+(7 \times \rm FWHM)^2\,]^{\,0.5}$ pixels; the outer radius
was chosen so that the annulus should contain a total of about 1000 pixels.
This choice was made to ensure a constant number of sky pixels independently
on the seeing conditions; otherwise, the accuracy of the mean local sky
background determination would vary depending on the number of pixels in the
sky annulus (the error of the mean over $N$ pixels is $N^{0.5}$ times smaller
than the error of the single pixel). Therefore, the uncertainty in flux measurements
due to the uncertainty of the sky level would get larger as the number of sky pixels
decreases.

A second-order polynomial correction for the presence of scattered
light\footnote{Scattered light affects both the science and the flat
field frames in the same manner. Being added to the flat field frame, it will
take part in a multiplicative way in the flat field correction; as a result the
sky background will be flat but the photometry will be dependent on the object position
on the CCD chip (see Manfroid, Selman \& Jones \cite{manfroid01}; Boyle et~al. \cite{boyle03}).
The presence of scattered light in the 2-m telescope of NAO was detected through
camera obscura experiments (H. Markov, private communication; see also Grundahl
\& S{\o}rensen \cite{grundahl96}).} in the telescope was applied according to
$$m_{\rm obs}-m_{\rm corr}=p\,(d/512.5)+q\,(d/512.5)^2,$$ where $m_{\rm obs}$
and $m_{\rm corr}$ are the observed and corrected instrumental magnitudes,
respectively, $p$ and $q$ are the polynomial coefficients, and $d$ is the
distance to the frame centre. The cluster standards observed during the
period 1997~-- 1999 were used to calculate the mean polynomial coefficients
of the correction.

The transformation of the total instrumental magnitudes to the standard system
was done following Harris, Fitzgerald \& Reed (\cite{harris81}) methodology. It
assumes determination of the extinction and photometric transformation coefficients
at once; the imaging of multi-star standard fields ensures the accurate
determination of the extinction coefficients even with two or three standard
observations during the night. The transformation equations read:
\begin{eqnarray}
u-U & = & c_{\rm \scriptstyle U}^{(0)}+c_{\rm \scriptstyle U}^{(1)}\,X+c_{\rm \scriptstyle U}^{(2)}\,(U-B) \nonumber \\
b-B & = & c_{\rm \scriptstyle B}^{(0)}+c_{\rm \scriptstyle B}^{(1)}\,X+c_{\rm \scriptstyle B}^{(2)}\,(B-V) \nonumber \\
v-V & = & c_{\rm \scriptstyle V}^{(0)}+c_{\rm \scriptstyle V}^{(1)}\,X+c_{\rm \scriptstyle V}^{(2)}\,(V-R) \nonumber \\
r-R & = & c_{\rm \scriptstyle R}^{(0)}+c_{\rm \scriptstyle R}^{(1)}\,X+c_{\rm \scriptstyle R}^{(2)}\,(V-R) \nonumber \\
i-I & = & c_{\rm \scriptstyle I}^{(0)}+c_{\rm \scriptstyle I}^{(1)}\,X+c_{\rm \scriptstyle I}^{(2)}\,(R-I), \nonumber
\end{eqnarray}
where the small and the capital letters denote the instrumental and the catalogue
magnitudes of the cluster standard stars, respectively; $X$ is the airmass, $c^{(0)}$
is the zero-point magnitude, $c^{(1)}$ is the extinction coefficient, and $c^{(2)}$ is
the colour coefficient. The transformation coefficients, $\vec{c}$, and their errors
were determined solving the linear set of equations $\rm \partial \chi^{2}(\vec{c})/\rm \partial \vec{c}=0$
for the total instrumental and catalogue magnitudes of the cluster standards employing
Gauss-Jordan algorithm; the weights applied were equal to the inverse sum of the square
errors of the instrumental and catalogue magnitudes. The transformation equations
were applied to the night data, so, the transformation coefficients were determined for
each observing night.

The total instrumental magnitudes of the comparison stars were transformed to the
standard Johnson-Cousins system inverting the above equations with the transformation
coefficients substituted. The uncertainties of the magnitudes were estimated by means
of the standard error propagation rules taking into account the errors of the instrumental
magnitudes as returned by $\rm \scriptstyle DAOPHOT$, the errors of the scattered light polynomial
correction coefficients, and the errors of the transformation coefficients. In all cases
of multi-epoch observations the calibrated magnitudes were weight-averaged; the errors
of the mean magnitudes were calculated taking the larger between (1) the error estimate
based on the individual magnitude errors and (2) the error estimate based on the scatter
of magnitudes involved in averaging about their weight-mean value (see Stetson, Bruntt \& Grundahl
\cite{stetson03}).

The calibrated apparent magnitudes of the comparison stars are presented in Table~\ref{naomag};
the equatorial coordinates have been obtained using $\rm \scriptstyle ALADIN$ web facility
(Bonnarel et~al. \cite{bonnarel00}). The comparison stars are denoted with capital letters in order
of increasing declination. We list the other designations of the stars having literature
calibrations as well as the corresponding papers in Table~\ref{naomag} in order to
facilitate the cross-identifications of the comparison stars. The finding charts are
presented in Fig.~\ref{fcharts}; they have been prepared using digitized $B$ or $R$
plates (depending on the quality) of the Second Palomar Observatory Sky Survey (POSS-II).

\begin{figure}
\begin{center}
\begin{minipage}{83mm}
\resizebox{\hsize}{!}{\includegraphics{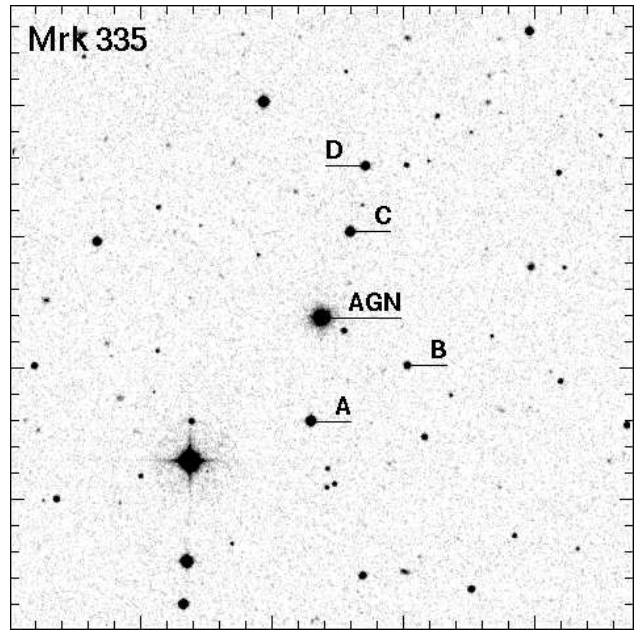}}
\end{minipage}
\hfill
\begin{minipage}{83mm}
\resizebox{\hsize}{!}{\includegraphics[bb= 70 301 522 778]{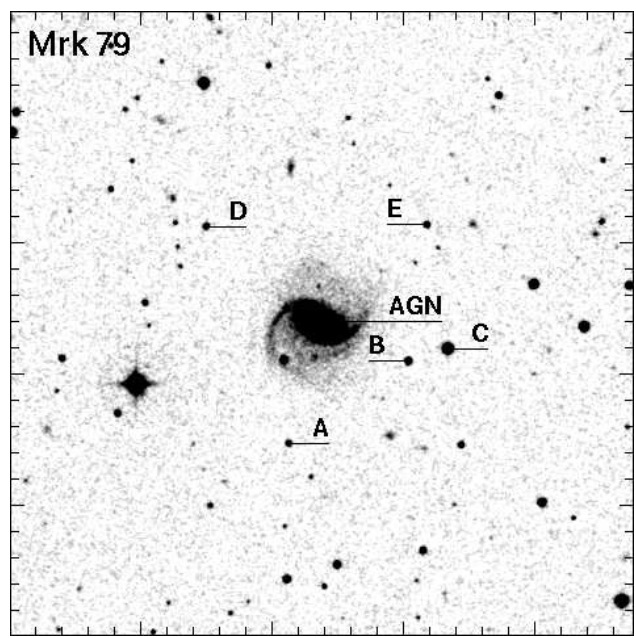}}
\caption{\label{fcharts}Finding charts for the comparison stars. The size of the charts is
$8 \times 8$ arcmin and the distance between the ticks is 20 arcsec. North is at
the top, East is to the left.}
\end{minipage}
\end{center}
\hfill
\end{figure}

\setcounter{figure}{0}

\begin{figure}
\begin{center}
\begin{minipage}{83mm}
\resizebox{\hsize}{!}{\includegraphics{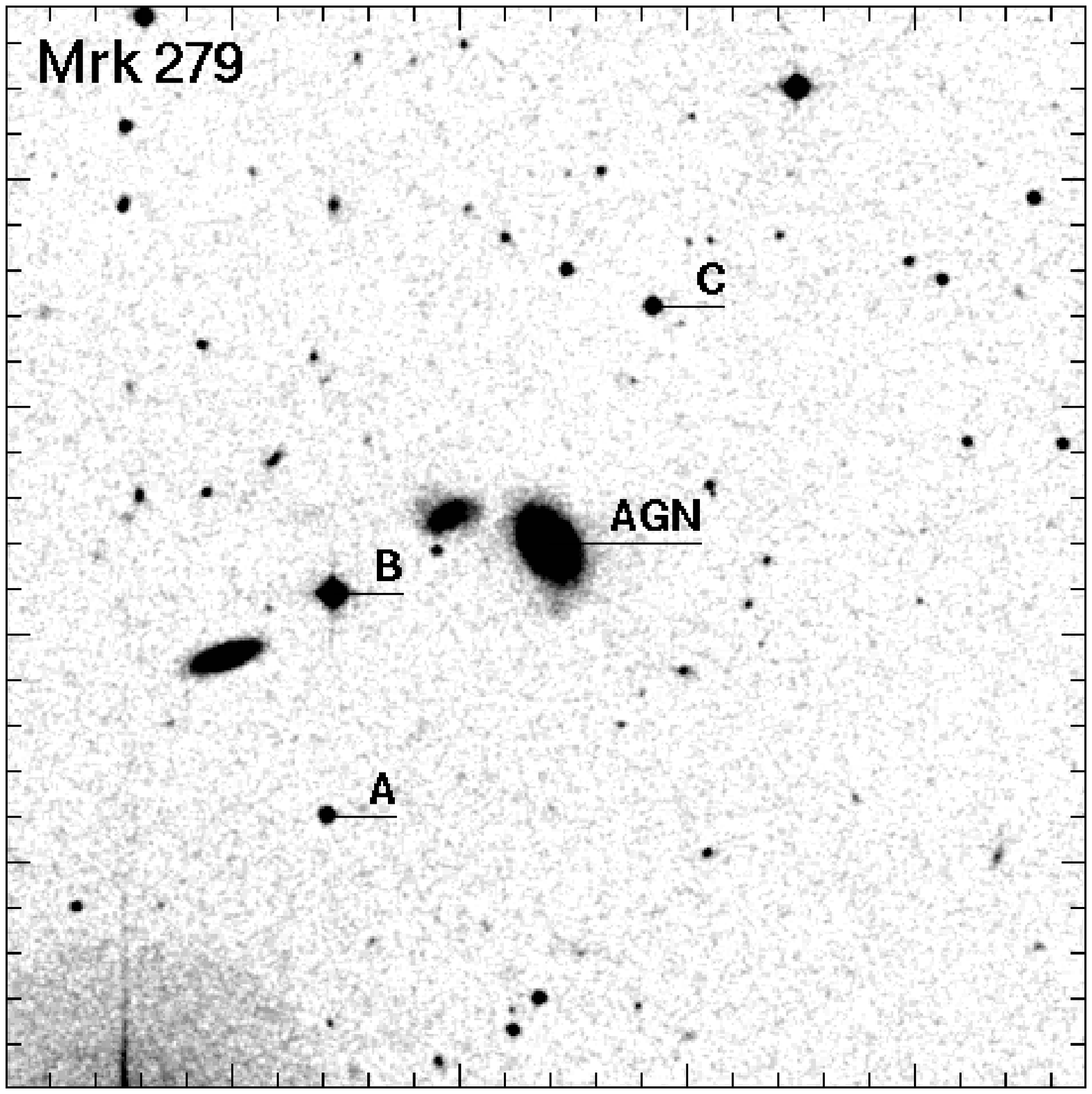}}
\end{minipage}
\hfill
\begin{minipage}{83mm}
\resizebox{\hsize}{!}{\includegraphics[bb= 70 301 522 778]{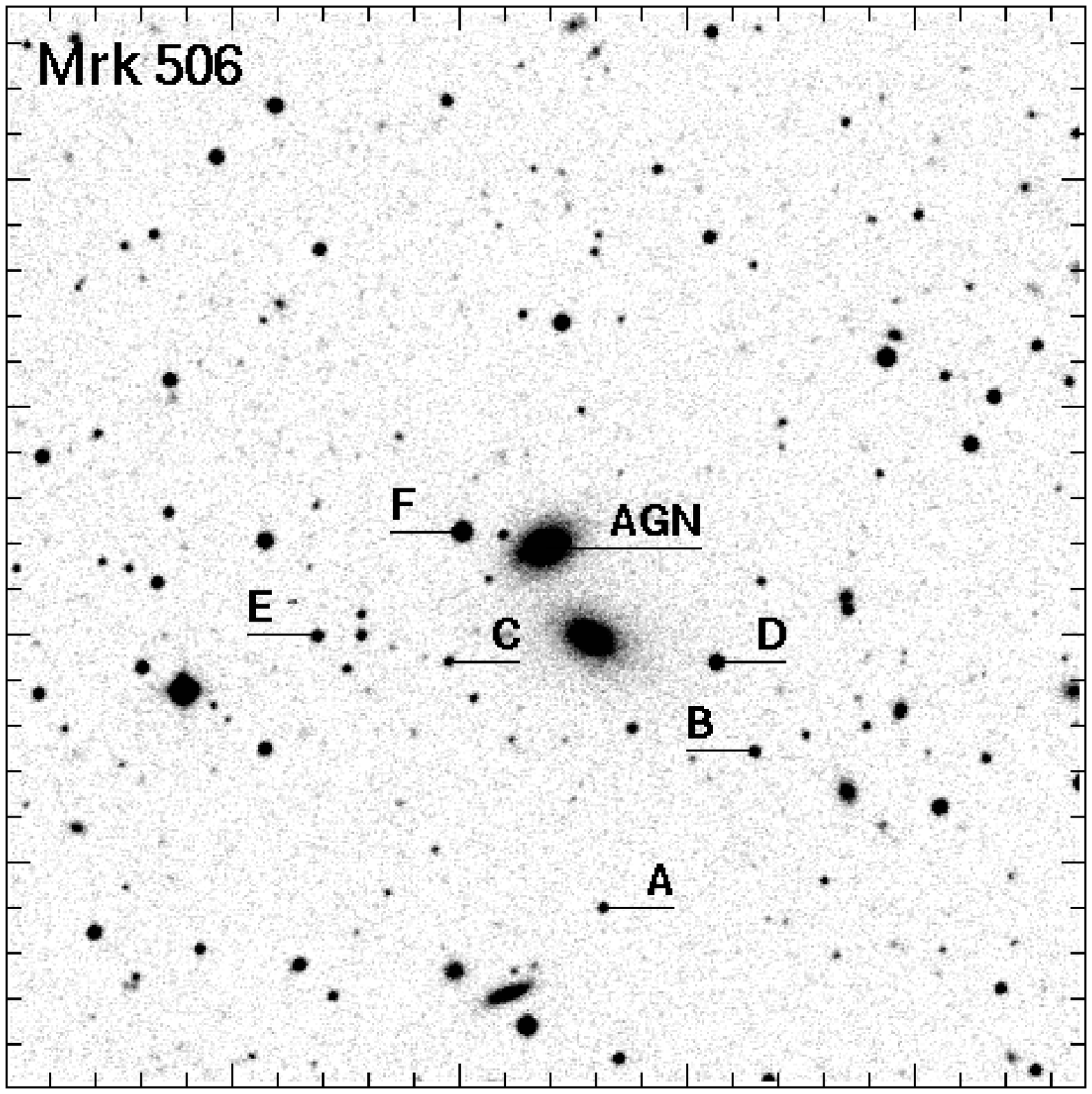}}
\caption{Continued.}
\end{minipage}
\end{center}
\hfill
\end{figure}

\setcounter{figure}{0}

\begin{figure}
\begin{center}
\begin{minipage}{83mm}
\resizebox{\hsize}{!}{\includegraphics{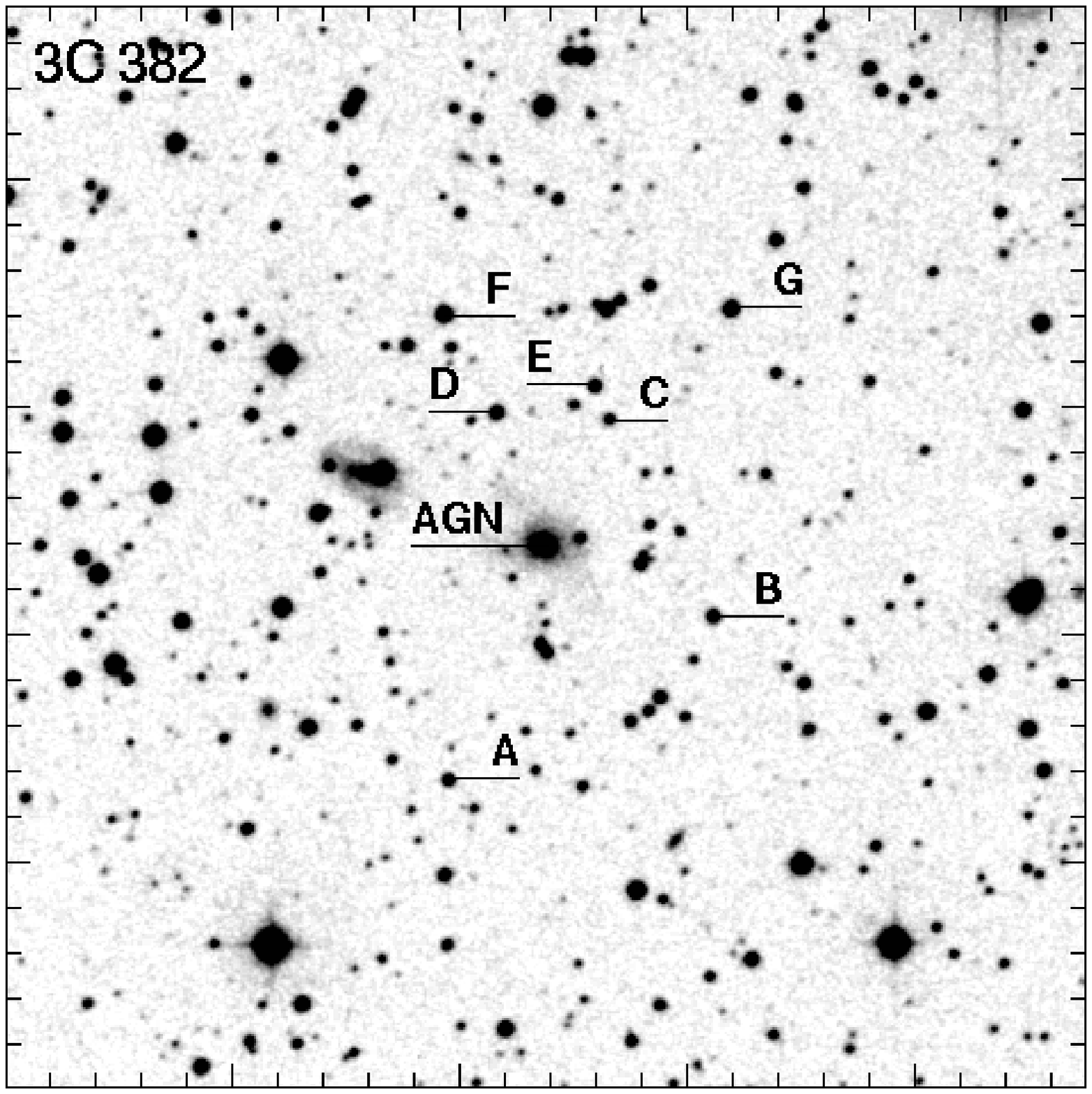}}
\end{minipage}
\hfill
\begin{minipage}{83mm}
\resizebox{\hsize}{!}{\includegraphics[bb= 70 301 522 778]{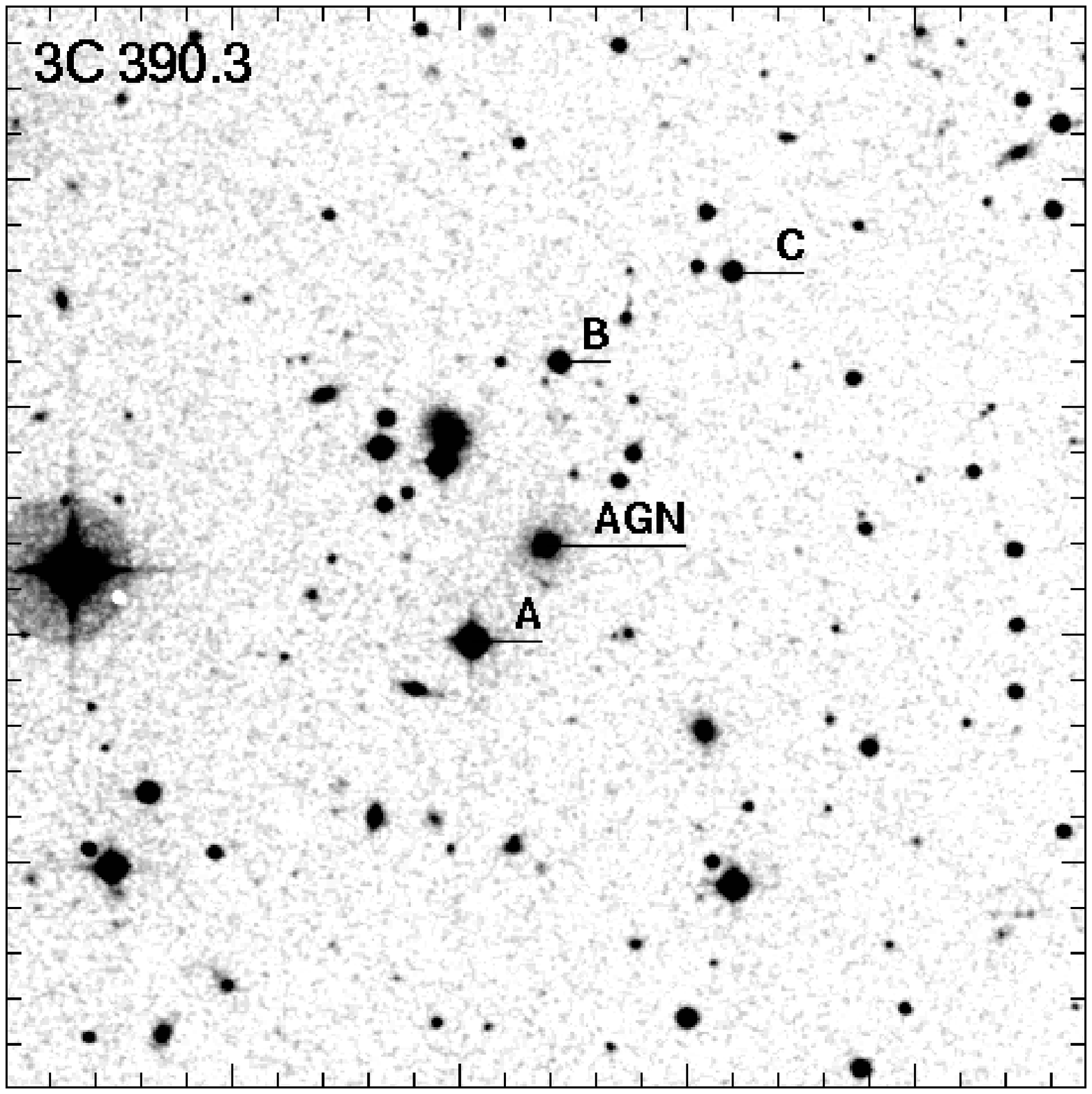}}
\caption{Continued.}
\end{minipage}
\end{center}
\hfill
\end{figure}

\setcounter{figure}{0}

\begin{figure}
\begin{center}
\begin{minipage}{83mm}
\resizebox{\hsize}{!}{\includegraphics{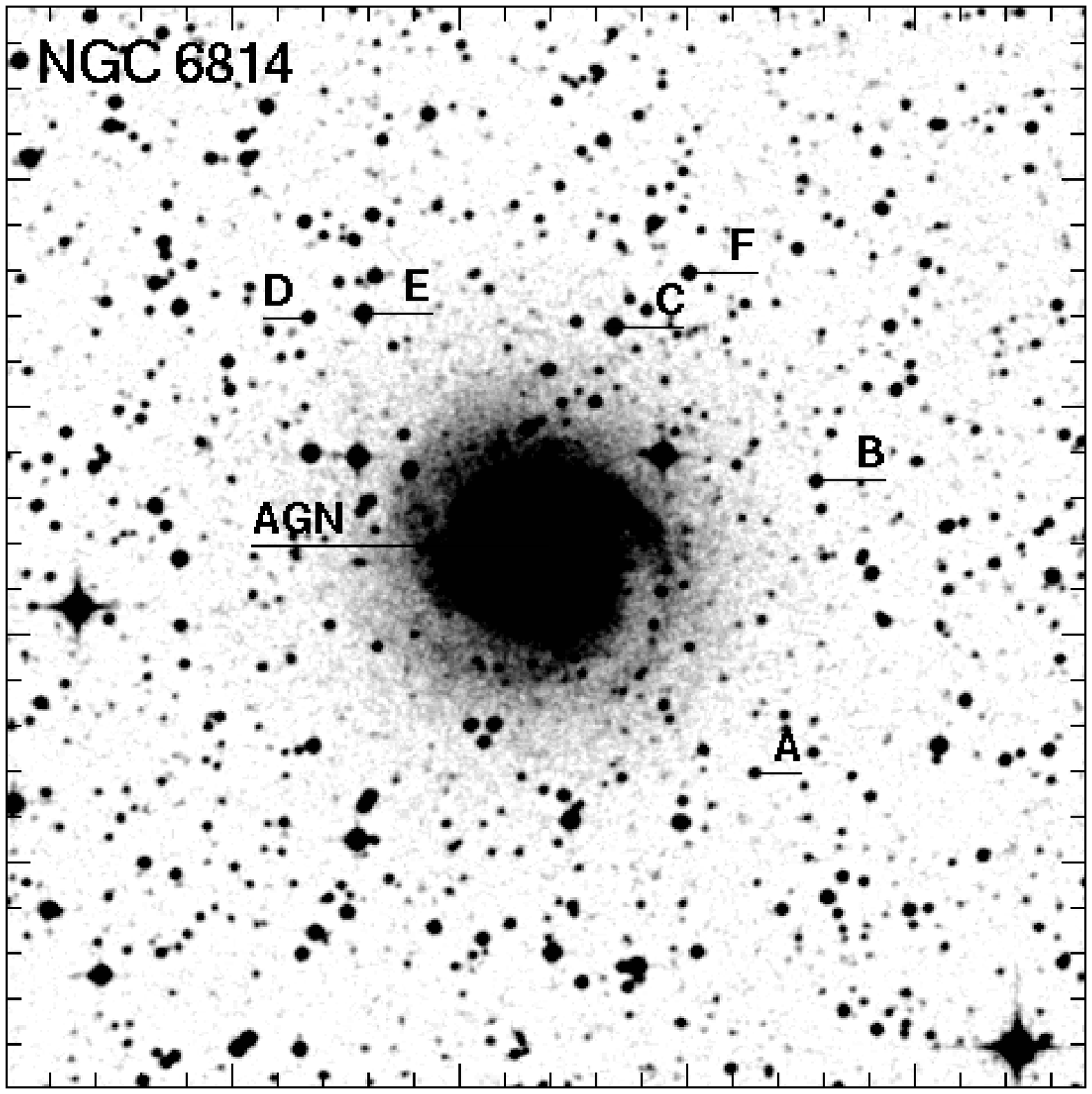}}
\end{minipage}
\hfill
\begin{minipage}{83mm}
\resizebox{\hsize}{!}{\includegraphics[bb= 70 301 522 778]{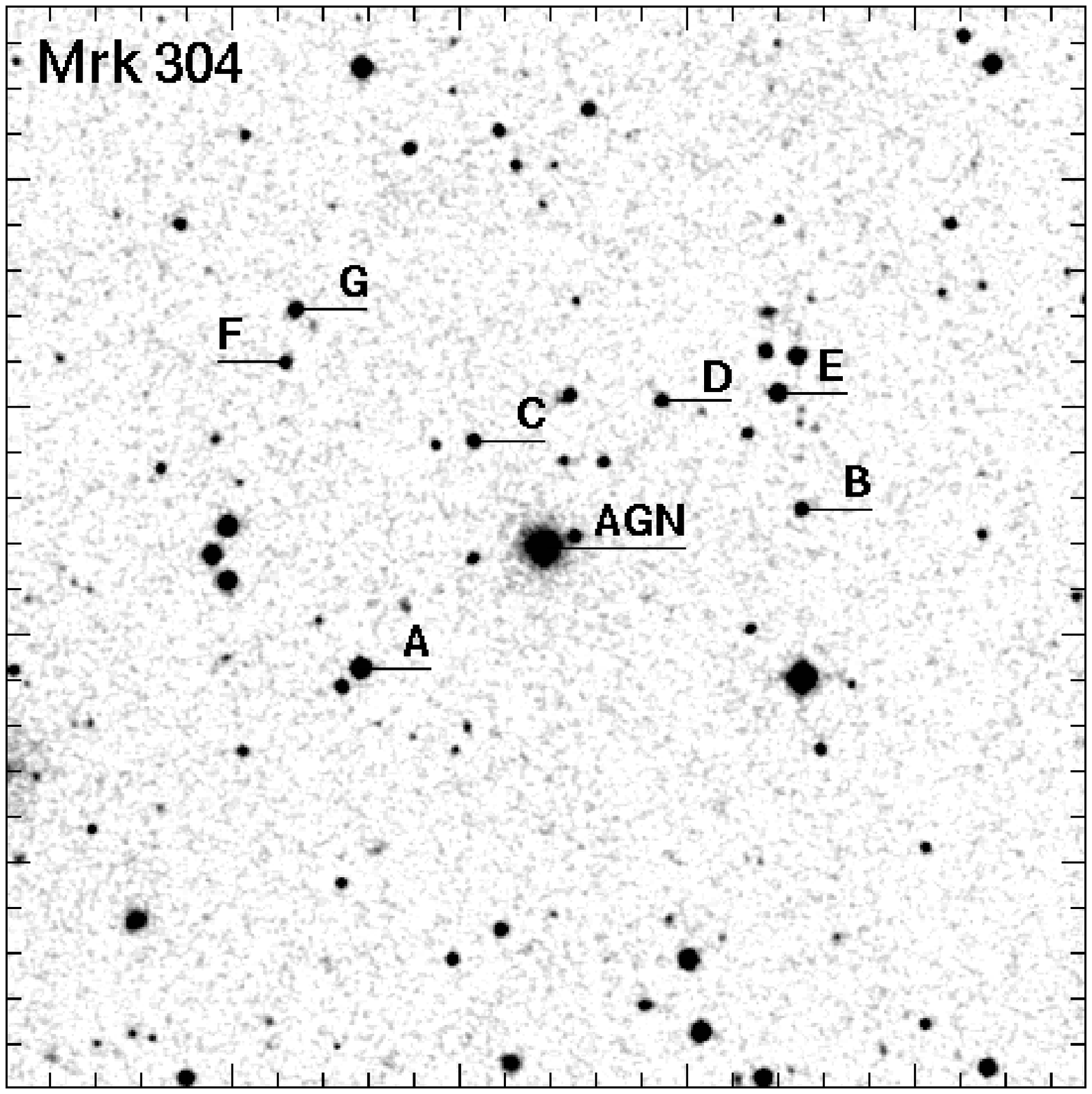}}
\caption{Continued.}
\end{minipage}
\end{center}
\hfill
\end{figure}

\setcounter{figure}{0}

\begin{figure}
\begin{center}
\begin{minipage}{83mm}
\resizebox{\hsize}{!}{\includegraphics{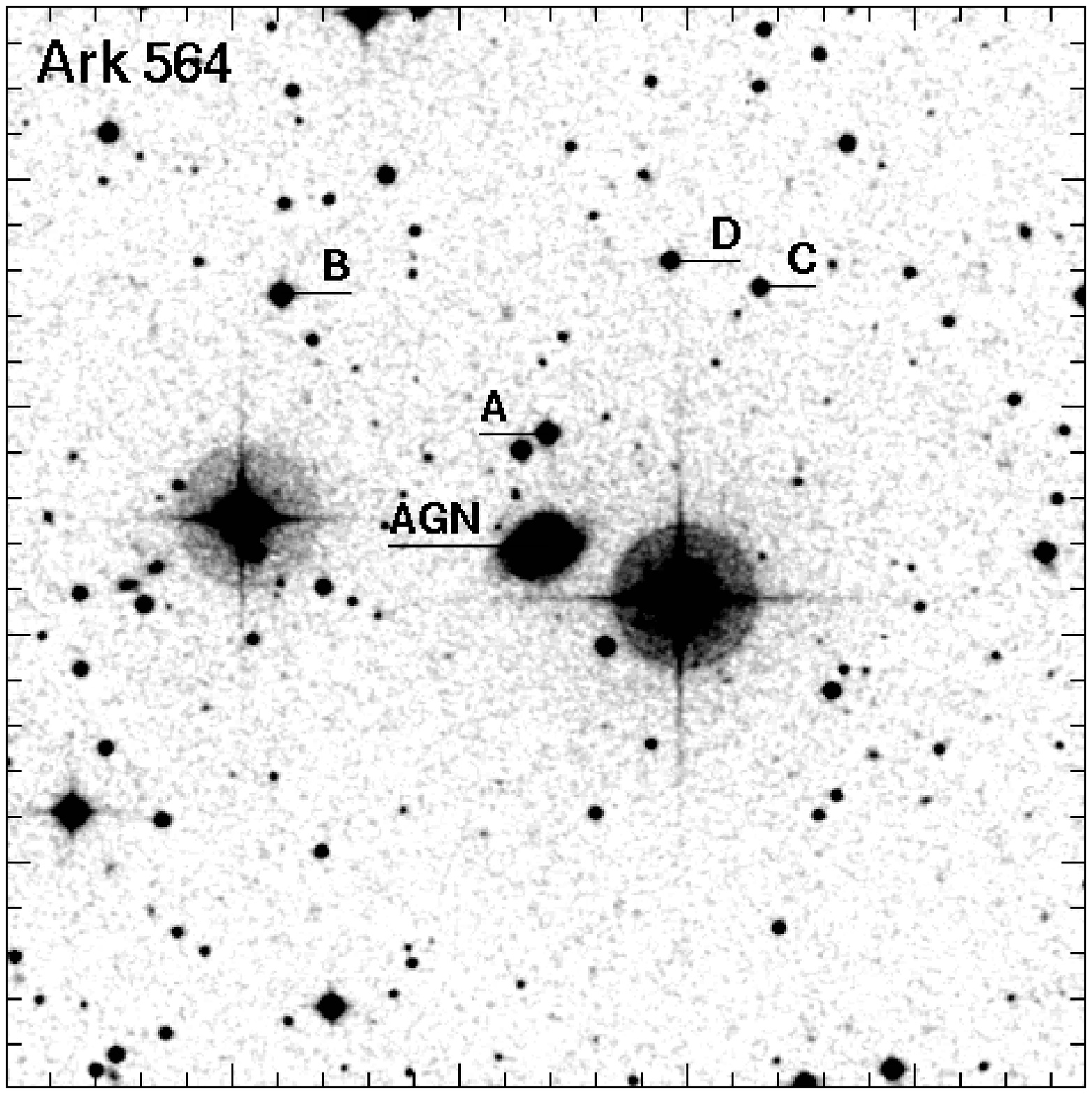}}
\end{minipage}
\hfill
\begin{minipage}{83mm}
\resizebox{\hsize}{!}{\includegraphics[bb= 70 301 522 778]{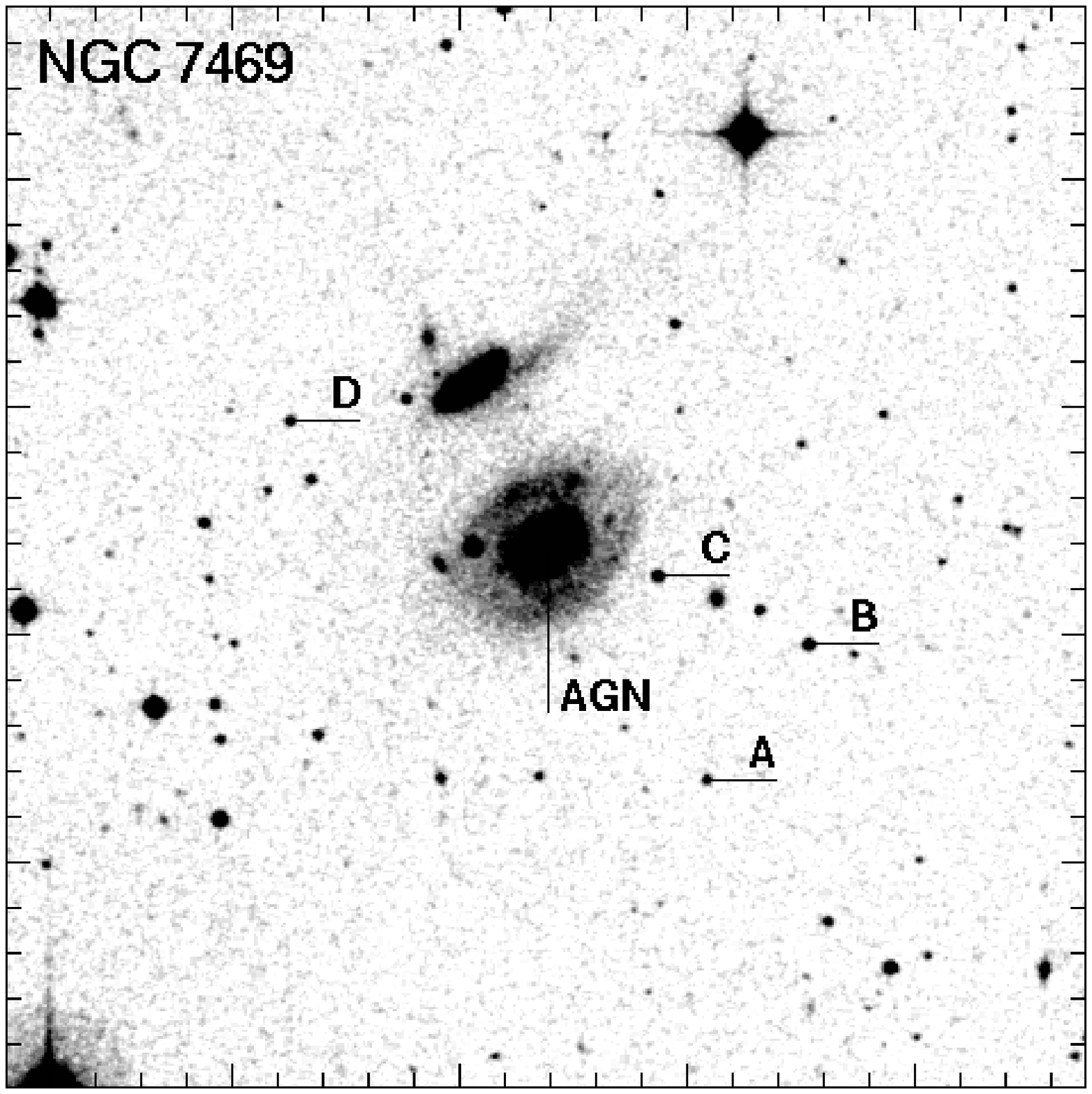}}
\caption{Continued.}
\end{minipage}
\end{center}
\hfill
\end{figure}

\section{General characteristics of the comparison stars}
\label{genchar}

The calibrated comparison stars cover $V$ band flux range from 11.7 to 18.2 mag with a
median value of 16.3 mag and $B-V$ colour index range from 0.4 to 1.6 mag with a median
value of 0.8 mag. The colour index range roughly corresponds to a spectral type interval
from F to M for main sequence stars. The lack of blue stars among the standards is
obvious~-- this could be considered as a disadvantage of the presented comparison stars
since Seyfert nuclei are blue objects.

The minimal, median, and maximal magnitude errors of the calibrated comparison stars
in respective order are the following: $U$ band~-- 0.04, 0.08, and 0.21 mag; $B$
band~-- 0.03, 0.04, and 0.09 mag; $V$ band~-- 0.02, 0.03, and 0.08 mag;
$R_{\rm \scriptstyle C}$ band~-- 0.02, 0.04, and 0.09 mag; $I_{\rm \scriptstyle C}$ band~-- 0.04,
0.06, and 0.11 mag. The largest errors are attributed to the $U$ band magnitudes (due
to the lower sensitivity of the CCD chip) and to the $I_{\rm \scriptstyle C}$ band
ones (due to the high sky background level and due to the presence of a fringe pattern).

We did several checks of the accuracy of our calibration and for the eventual presence of
variable stars in the calibrated comparison sequences.

Firstly, all comparison stars were queried in $\rm \scriptstyle SIMBAD$ astronomical database
for eventual coincidence with known variable stars. The results were negative for
a search radius of 30 arcsec around the star positions.

Secondly, an internal check for variability was done for the stars calibrated for the first time.
This was done by means of a multiband differential photometry of the comparison stars relative
to the brightest one in the corresponding field. We complemented the photometric night data with data
from non-calibrated nights (see Sect.~\ref{intro}) and with data obtained during 2007 June observing
run\footnote{We observed Mrk~506, 3C~382, and Mrk~304 fields with the 2-m telescope of NAO in a
single night during this run. These fields were imaged through Johnson-Cousins $BVR_{\rm \scriptstyle C}I_{\rm \scriptstyle C}$
filters using $1340 \times 1300$ Princeton Instruments VersArray:1300B model CCD camera that has
EEV (Marconi) CCD36--40 chip with $20\,\mu\rm m$ square pixel ($0.258\,\rm arcsec\,\rm px^{-1}$
scale factor). Observations and data reduction were performed in a manner similar to that described
in Sect.~\ref{obsred}. No calibration was performed due to the unstable weather conditions.} in order
to increase the significance of the internal check; the number of nights used in the internal check
for variability of the newly calibrated comparison stars is specified in Table~\ref{gals}. We found
the mean absolute deviation about the median instrumental magnitude difference to be compatible with
or less than the median error of the calibrated magnitudes for each passband and star light curve.
A few exceptions were found that could be attributed to single bad measurements rather than to
variability. Note that an eventual large difference between the individual magnitude estimates
will result~-- after the weight-averaging~-- in a large error in the final magnitude. Based
on the above considerations we could conclude that the comparison stars calibrated for the
first time could be assumed to be non-variable down to amplitudes compatible with the
uncertainties of their calibrated magnitudes.

\begin{figure}[t]
\resizebox{\hsize}{!}{\includegraphics{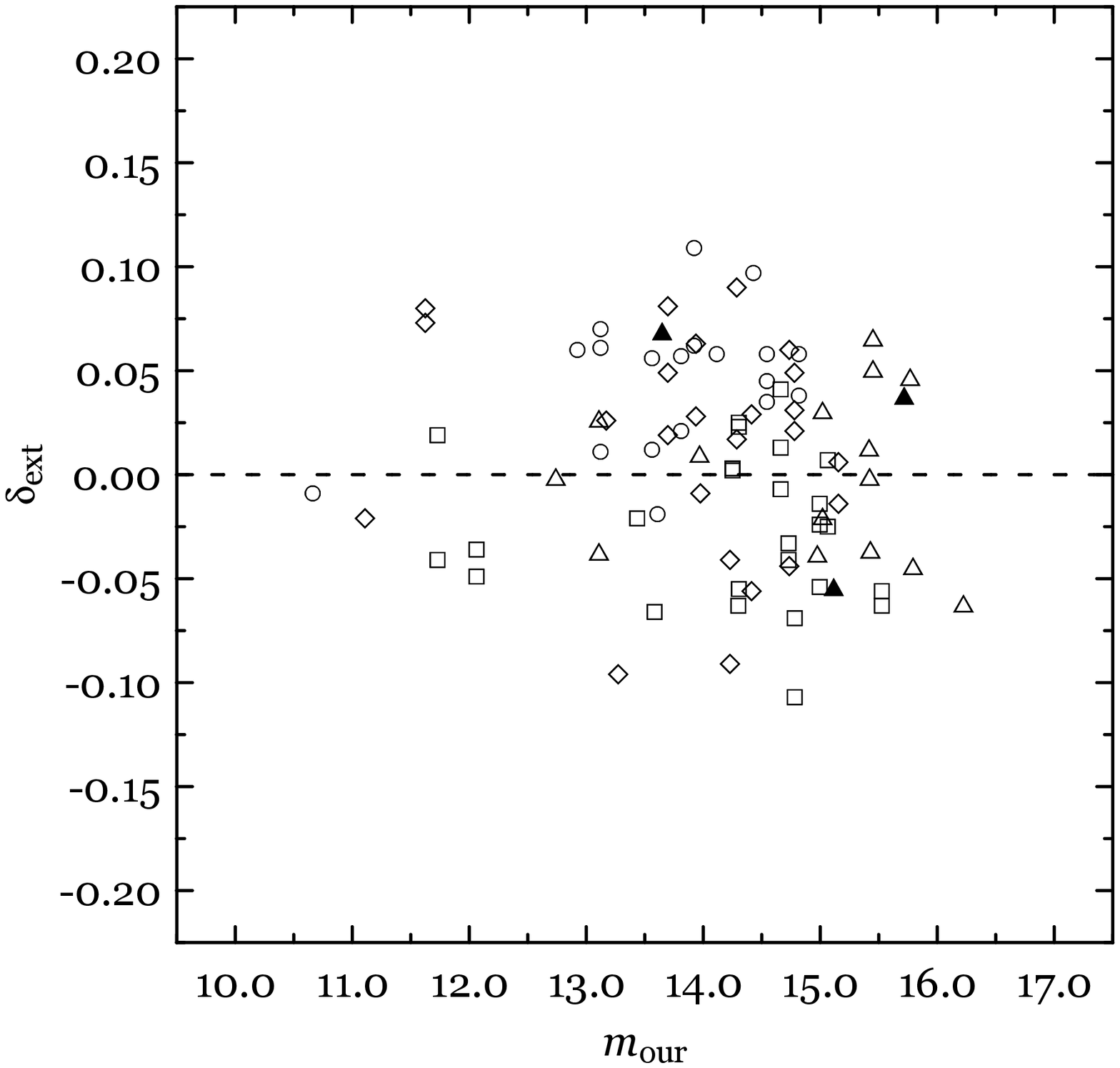}}
\caption{\label{ecomp1}An external check of our calibration: the values of $\delta_{\rm ext}$
are plotted against $m_{\rm our}$. Different passbands are marked
as follows: $U$~-- filled triangles, $B$~-- open triangles, $V$~-- squares,
$R_{\rm \scriptstyle C}$~-- diamonds, and $I_{\rm \scriptstyle C}$~-- circles.
Individual objects are not specified for
the sake of clarity.}
\end{figure}

\begin{figure}[t]
\resizebox{\hsize}{!}{\includegraphics{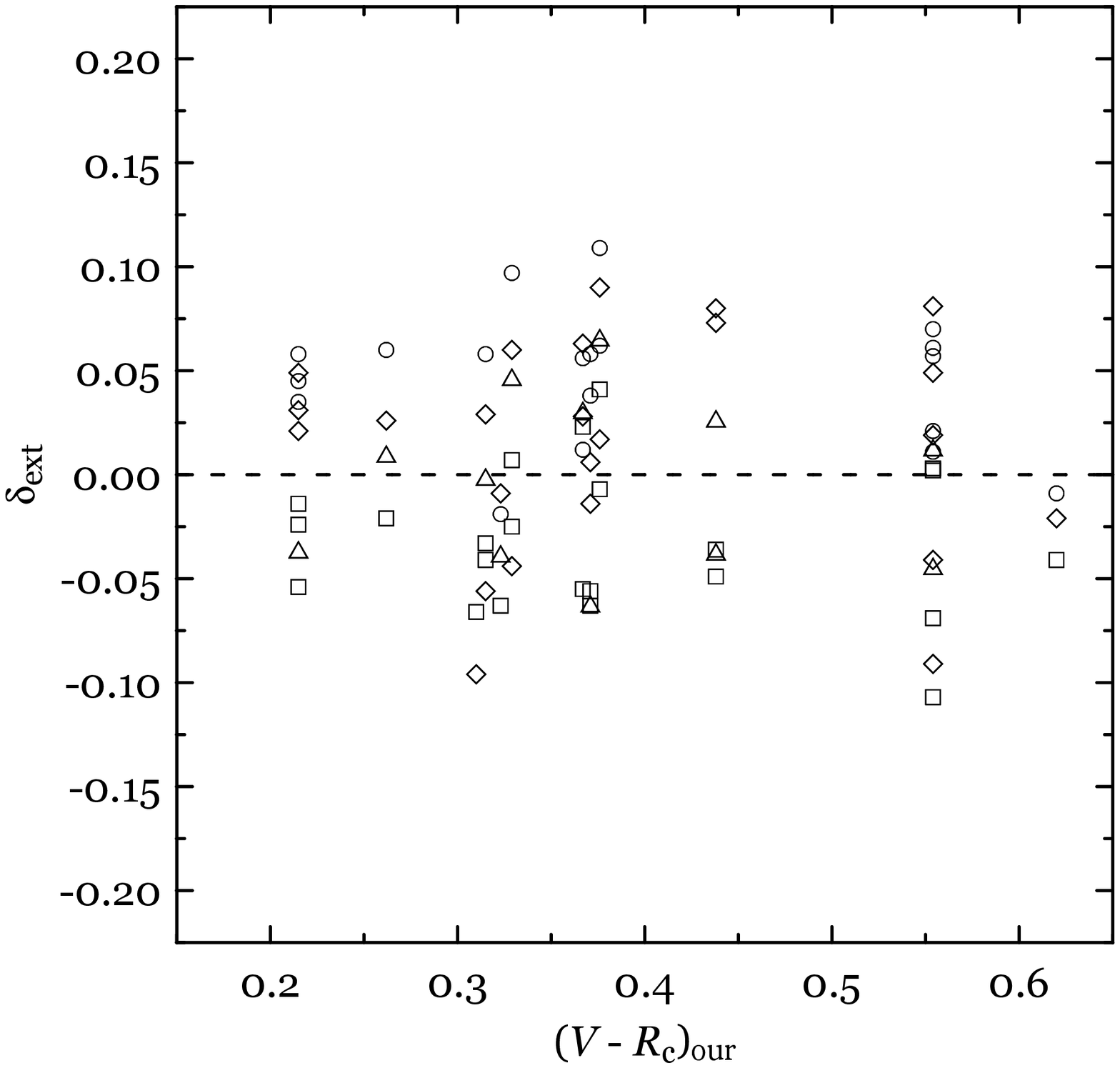}}
\caption{\label{ecomp2}An external check of our calibration: the values of $\delta_{\rm ext}$
are plotted against $(V-R_{\rm \scriptscriptstyle C})_{\rm our}$. Different passbands are marked
as in Fig.~\ref{ecomp1}.}
\end{figure}

And finally, we considered an external check of our calibration in the cases when literature results
were found; note that {\em Ic1} and {\em Ic2} magnitudes published by Doroshenko et~al. (\cite{doroshenko05a},
\cite{doroshenko05b}) were weight-averaged before their usage in the external check. In Fig.~\ref{ecomp1}
the differences between our magnitudes and the magnitudes calibrated by other authors,
$\delta_{\rm ext}=m_{\rm our}-m_{\rm other}$, are plotted against our magnitudes, $m_{\rm our}$,
listed in Table~\ref{naomag}. In Fig.~\ref{ecomp2} we plot $\delta_{\rm ext}$ against the
$V-R_{\rm \scriptstyle C}$ colour index of the comparison stars. We found the median value
of the quantity $|\delta_{\rm ext}|/(\sigma_{\rm our}^2+\sigma_{\rm other}^2)^{0.5}$ (where
$\sigma_{\rm our}$ and $\sigma_{\rm other}$ are the errors of our magnitudes and of the
literature ones, respectively) to be less than unity for all passbands. These considerations
suggest that there are no significant systematic deviations of our calibration from the
previously published results depending on the passband, colour index or magnitude. So,
we could conclude that the differences between our magnitudes and the literature ones are
caused by measurement and calibration uncertainties and, therefore, our calibrations agree
with the literature ones to within the errors. This could also be considered as evidence for
non-variability of the re-calibrated stars compared to their previous calibrations down to
amplitudes compatible with the magnitude errors.

\section{Comments on the individual fields}
\label{comm}

{\em Mrk~335.} Calibrations of stars in this field were done by Curry et~al. (\cite{curry98})~-- three stars in
the $VR_{\rm \scriptstyle C}I_{\rm \scriptstyle C}$ bands, Bachev et~al. (\cite{bachev00})~-- four stars in
the $VR_{\rm \scriptstyle C}I_{\rm \scriptstyle C}$ bands, and Doroshenko et~al. (\cite{doroshenko05a})~--
seven stars in the $BVR_{\rm \scriptstyle C}I_{\rm \scriptstyle C}$ bands. There are two stars
calibrated by us, A and C, that have been measured by the above authors as well; our calibration confirms
the literature results to within the errors. We added two new comparison stars, thus extending
the sequence to fainter magnitudes compared to the previous calibrations~-- our star B is the
faintest one calibrated in this field so far. We calibrated $U$ band magnitudes of stars in this
field for the first time.

{\em Mrk~79}. Calibration of stars in this field was done by Doroshenko et~al. (\cite{doroshenko05a})~-- six stars in
the $BVR_{\rm \scriptstyle C}I_{\rm \scriptstyle C}$ bands. We have only one common star
with them~-- our star C is their star 4. The other stars calibrated by  Doroshenko et~al. (\cite{doroshenko05a}) are
outside our field of view. We added four new comparison stars that are closer to the galaxy and
fainter than the stars previously calibrated. We also calibrated $U$ band magnitudes of stars in
this field for the first time. The magnitudes of star C are single epoch ones because the star was
outside our field of view at the second epoch. This field was also included in the paper
of Curry et~al. (\cite{curry98}), but no calibrated magnitudes were reported by them.

{\em Mrk~279}. Calibrations of stars in this field were done by Bachev et~al. (\cite{bachev00})~-- four stars in the
$VR_{\rm \scriptstyle C}I_{\rm \scriptstyle C}$ bands\footnote{The $B$ band magnitude
of star A of Bachev et~al. (\cite{bachev00}) is presented in Bachev \& Strigachev (\cite{bachev04}).},
and  Doroshenko et~al. (\cite{doroshenko05b})~--
four stars in the $BVR_{\rm \scriptstyle C}I_{\rm \scriptstyle C}$ bands. We were not
able to add new comparison stars, but confirmed the previous calibrations of some of the stars
(see Table~\ref{naomag}) to within the errors and extended the passband coverage of the
calibrated magnitudes towards the $U$ band. We were not able to calibrate star B in the
$R_{\rm \scriptstyle C}I_{\rm \scriptstyle C}$ bands because it was
saturated in the corresponding frames.

{\em Mrk~506}. No other $UBVR_{\rm \scriptstyle C}I_{\rm \scriptstyle C}$
calibrations of this field exist to our knowledge.

{\em 3C~382}. No other $UBVR_{\rm \scriptstyle C}I_{\rm \scriptstyle C}$ calibrations
of this field exist to our knowledge. The second epoch $B$ band frames of this field were of low quality
due to scattered light and the corresponding magnitudes were discarded because they showed large
differences compared to the first epoch measurements, i.e. no weight-averaging was performed in
this case.

{\em 3C~390.3}. Calibrations of stars in this field were done by Penston et~al. (\cite{penston71})~-- three stars in the
$UBV$ bands, Curry et~al. (\cite{curry98})~-- three stars in the $VR_{\rm \scriptstyle C}I_{\rm \scriptstyle C}$
bands, and Doroshenko et~al. (\cite{doroshenko05b})~-- eleven stars in the $BVR_{\rm \scriptstyle C}I_{\rm \scriptstyle C}$
bands. We re-calibrated some of the stars (see Table~\ref{naomag}) measured by the above authors
confirming their results to within the errors over all passbands. We were not able to obtain two-epoch
data for star A in the $R_{\rm \scriptstyle C}I_{\rm \scriptstyle C}$ bands because it was
saturated in the corresponding second epoch frames.

{\em NGC~6814}. Calibration of stars in this field was done by Doroshenko et~al. (\cite{doroshenko05b})~--
six stars in the $BVR_{\rm \scriptstyle C}I_{\rm \scriptstyle C}$ bands. We have no
common stars with them because their stars 3, 6, 8, and 9 are outside our field of view,
whereas stars 1 and 2 are saturated in our frames. We added six new stars to the field
of the object, fainter than previously calibrated standards; $U$ band magnitudes
were also calibrated by us. Star A was not detected in our $U$ band frames, so, we could
not obtain $U$ band magnitude for it. We excluded the second epoch $B$ magnitudes
and the first epoch $I_{\rm \scriptstyle C}$ ones from weight-averaging due to the
large differences compared to the other epoch measurements.

{\em Mrk~304}. No other $UBVR_{\rm \scriptstyle C}I_{\rm \scriptstyle C}$
calibrations of this field exist to our knowledge.

{\em Ark~564}. Calibrations of stars in this field were done by Shemmer et~al. (\cite{shemmer01})~-- three stars in the
$BVR_{\rm \scriptstyle C}I_{\rm \scriptstyle C}$ bands, and Doroshenko et~al. (\cite{doroshenko05b})~--
twelve stars in the $BVR_{\rm \scriptstyle C}I_{\rm \scriptstyle C}$ bands. We did not add
new comparison stars to this field, but we re-calibrated some of the stars (see Table~\ref{naomag})
measured by the above authors and added $U$ band magnitudes. We confirmed the literature results
to within the errors with the following exception: we found the $BI_{\rm \scriptstyle C}$ magnitude
differences for our stars A, C, and D compared to stars 1, 2, and 3 of Shemmer et~al. (\cite{shemmer01}) in the field
of Ark~564 to be $\delta_{\rm ext}$~$\approx$~$-0.6\,\rm mag$; differences of the same order were found
by Doroshenko et~al. (\cite{doroshenko05b}). On the other hand, our $VR_{\rm \scriptstyle C}$ magnitudes are
in good agreement with both sets of literature results. Therefore, there is some kind of
error or misprint concerning the $BI_{\rm \scriptstyle C}$ magnitudes of stars 1, 2,
and 3 presented by Shemmer et~al. (\cite{shemmer01}). The magnitude differences corresponding to these
deviated magnitudes were not plotted in Fig.~\ref{ecomp1} and in Fig.~\ref{ecomp2}. Doroshenko et~al. (\cite{doroshenko05b})
claimed that their star 12 may be a low-amplitude variable (see Table~\ref{naomag} for the other
designations of this star). We cannot confirm or reject this finding because our magnitude
errors are larger than the variability amplitudes estimated by Doroshenko et~al. (\cite{doroshenko05b}).

{\em NGC~7469}. Calibrations of stars in this field were done by Penston et~al. (\cite{penston71})~-- five stars in the
$UBV$ bands, and Doroshenko et~al. (\cite{doroshenko05b})~-- nine stars in the $BVR_{\rm \scriptstyle C}I_{\rm \scriptstyle C}$
bands. We have no common stars with both groups because their stars are outside our field of view.
We added four new stars to the field of the object, closer to the galaxy and fainter than
previously calibrated standards.

\section{Summary}
\label{summ}

We have calibrated Johnson-Cousins $UBVR_{\rm \scriptstyle C}I_{\rm \scriptstyle C}$
magnitudes of 49 stars in the fields of the Seyfert galaxies Mrk~335, Mrk~79, Mrk~279,
Mrk~506, 3C~382, 3C~390.3, NGC~6814, Mrk~304, Ark~564, and NGC~7469 in order to facilitate
photometric monitoring of these objects; 36 stars have been measured for the first time.
Our magnitudes are in good agreement with the published ones and we have found no signs
of variability of the calibrated comparison stars at least with amplitudes
down to the estimated magnitude errors.

We have calibrated in Johnson-Cousins $UBVR_{\rm \scriptstyle C}I_{\rm \scriptstyle C}$
system the comparison stars in three of the fields, Mrk~506, 3C~382, and Mrk~304, for the first
time (to our knowledge) and we have improved the existing standard sequences in the other fields.
New comparison stars have been added to the fields of Mrk~335 (two stars), Mrk~79 (four stars), NGC~6814
(six stars), and NGC~7469 (four stars)~-- most of the newly added stars are fainter and are situated
closer to the Seyfert galaxies compared to the existing standards. The passband coverage
of the comparison sequences in the fields of Mrk~335, Mrk~79, Mrk~279, NGC~6814, and Ark~564
have been complemented with the $U$ band.

Future observations of the presented comparison sequences are welcome in order
to improve them further.

\acknowledgements
The authors are thankful to the anonymous referee whose constructive suggestions
and criticism helped us to improve this paper.

This research has made use of the NASA/IPAC Extragalactic Database (NED) which is operated by the
Jet Propulsion Laboratory, California Institute of Technology, under contract with the National
Aeronautics and Space Administration.

This research has made use of the SIMBAD database, operated at CDS, Strasbourg, France.

The European Southern Observatory Munich Image Data Analysis System ($\rm \scriptstyle ESO-MIDAS$)
is developed and maintained by the European Southern Observatory.

The Digitized Sky Survey was produced at the Space Telescope Science Institute under U.S.
Government grant NAG W-2166. The images of these surveys are based on photographic data
obtained using the Oschin Schmidt Telescope on Palomar Mountain and the UK Schmidt Telescope.
The plates were processed into the present compressed digital form with the permission of
these institutions.

The Second Palomar Observatory Sky Survey (POSS-II) was made by the California Institute of
Technology with funds from the National Science Foundation, the National Aeronautics and Space
Administration, the National Geographic Society, the Sloan Foundation, the Samuel Oschin
Foundation, and the Eastman Kodak Corporation.

We also acknowledge the support by UNESCO-ROSTE for the regional collaboration.

\end{document}